\documentclass[twocolumn,showpacs,preprintnumbers,amsmath,amssymb,superscriptaddress]{revtex4}
\usepackage[bookmarks=false]{hyperref}
\usepackage{amssymb}
\usepackage{amsmath}
\usepackage{graphicx}
\usepackage{dcolumn}
\usepackage{bm}
\usepackage{epstopdf}
\usepackage{times}
\usepackage{color}
\begin{document}
\preprint{}
%
\title{Large magntocaloric effect and 3D Ising critical behaviour in Gd$_2$Cu$_2$In}
%
%
%
\author{K. Ramesh Kumar*}
\affiliation{Highly Correlated Matter Research Group, Department of Physics, P. O. Box 524, University of Johannesburg, Auckland Park 2006, South Africa.}
\email{kraamesh57@gmail.com}
\author{Harikrishnan S. Nair*}
\affiliation{Department of Physics, University of Texas at El Paso, 500 W University Ave, El Paso, TX 79968, USA}
\author{B. N. Sahu}
\affiliation{Highly Correlated Matter Research Group, Department of Physics, P. O. Box 524, University of Johannesburg, Auckland Park 2006, South Africa.}
\author{Sindisiwe  Xhakaza}
\affiliation{Highly Correlated Matter Research Group, Department of Physics, P. O. Box 524, University of Johannesburg, Auckland Park 2006, South Africa.}
\author{Andr\'{e} M. Strydom}
\affiliation{Highly Correlated Matter Research Group, Department of Physics, P. O. Box 524, University of Johannesburg, Auckland Park 2006, South Africa.}
%
%
\begin{abstract}
The ternary intermetallic compound Gd$_2$Cu$_2$In crystallizes in Mo$_2$Fe$_2$B type structure with the space group $P4/mbm$ and we study critical behaviour and magnetocaloric effect near the ferromagnetic transition ($T_C$ $\approx$ 94 K) using the magnetic and heat capacity measurements. The maximum entropy change ($\Delta S_m$) and adiabatic temperature change ($\Delta T_{ad}$) for the field value of 7 T were  observed to be 13.8 J/kg.K and 6.5 K respectively. We have employed modified Arrott plot (MAP), Kouvel-Fisher (KF) procedures to estimate the critical exponents near the FM-PM phase transition. Critical exponents $\beta$ = 0.312(2), $\gamma$ = 1.080(5) are self-consistently estimated from the non-linear fitting. The $\beta$ value is close to the three dimensional (3D) Ising model where as $\gamma$ and $\delta$ values are close to mean field model. The estimated critical exponents for  Gd$_2$Cu$_2$In suggest that the system may belong to different universal class. All the three critical exponent obey Widom scale and collapse the scaled magnetic isotherms into two distinct branches below and above $T_C$ in accordance with single scaling equation. Specific heat measurements show a $\lambda$ type peak near 94 K confirming the bulk magnetic ordering. The data near $T_C$ was fitted using the non-linear function  $C_{P} = B + C\epsilon + A^{\pm}|\epsilon|^{-\alpha}(1 + E^{\pm} |\epsilon|^{0.5})$ between -0.025$<\epsilon<$0.025 which yielded the fourth critical exponent $\alpha$ value to be 0.11 (3). The value indicates possible 3D-Ising behavior where Gd$^{3+}$ moments arranged uniaxially along long tetragonel axis 'c' as reported in literature.
\end{abstract}
\maketitle
\section{\label{sec:level1}Introduction}

Magnetic materials with large magnetocaloric effect have attracted considerable attention due to their potential applications in magneto-refrigerant and eco-friendly cooling industry \cite{tishin2003magnetocaloric,Bruck,gschneidnerjr2005recent}. Among many possible materials, RTX (R = Rare-earth; T = Transition metal X = p block element), RCo$_2$ (R = Er, Dy, Ho) based alloys \cite{Singh_Rco2}  Gd$_5$Si$_2$Ge$_2$ and related compounds \cite{GdSi2Ge2}; Ni-Mn-X based Heusler alloys, RMnO$_3$ (R = Lanthanide) manganites \cite{Phan}; MnAs compounds \cite{Bruck}; and LaFe$_{13-x}$Si$_x$ based intermetallic compounds \cite{LaFe13Si} have drawn much focus due to giant magnetocaloric effect and large relative cooling power (RCP). The compounds R$_2$Cu$_2$In crystallize in the tetragonal structure which can be derived from  U$_3$Si$_2$ type structure compound. The R and indium atoms occupy the U sites whereas transition metal atoms are located at the Si site \cite{Canfield,Hulliger}. Magnetic and physical properties of Gd$_2$Cu$_2$In were reported by Fisher {\it et al.} on single crystalline sample and they found the system orders magnetically below 85 K where Gd moments are aligned along c-axis \cite{Canfield}. Gondek {\it et al.} have studied the magnetic and transport properties of Co substituted Gd$_2$Cu$_2$In samples and observed enhancement of ferromagnetic correlations with Co substitution \cite{Gondek}. Low-temperature magnetocaloric effect has been reported in R$_2$Cu$_2$X (R = Er, Tm, Ho and Dy X = Cd, In) systems recently many researchers \cite{Zhang_SR,ErCuIn,HoCuCd,GdCuCd,HoCuIn,DyCuIn}. The system Dy$_2$Cu$_2$In shows two successive phase transition at 49.5 K and 19.5 K belonging to a FM-PM transition and spin reorientation transition respectively. Dy$_2$Cu$_2$In showed large magnetocaloric effect (16.5 J/kg.K) with RCP value of 617 J/Kg \cite{ErCuIn} at 7 T. The maximum values of magnetic entropy change for Ho$_2$(Cu,Au)$_2$In are 21.9 (Cu) and 15.8 J/kg K (Au) under a field change of 0–7 T \cite{HoCuIn}.  Critical exponents analysis is an effective tool for understanding the nature of magnetic transitions and to study the intrinsic nature of the ferro-magnetism in alloyed compounds, amorphous ferromagnet, Rare-earth maganites \cite{GdNiAl2,LSMO,RCo3B2,Amorphous}. In this manuscript we report the magneto thermal and magnetocaloric effect on in Gd$_2$Cu$_2$In compound. Critical behavior and scaling analysis are examined using the conventional MAP, KF and critical scaling methods. 
\section{Experimental Details}
Polycrystalline sample was synthesized by arc-melting the stoichiometric mixtures of the high purity (99.9 wt \%  or better) elements under argon atmosphere in a Edmund Bühler (GmbH) arc-melting furnace. The sample was heat treated under high vacuum for one week at 800 $^{\circ}$C followed by ice water quenching. Crystal structure analysis is done using powder x-ray diffraction method and Rietveld refinement using GSAS software package \cite{EXPGUI,GSAS}. The isothermal magnetization $M(H)$ data were taken with close temperature intervals employing a commercial SQUID magnetometer (Quantum Design, Inc, San Diego). The specific heat ($C_P$) measurements were carried out on the Quantum Design PPMS (Quantum Design, Inc, Sand Diego) by the two-$\tau$ relaxation method under quasi-adiabatic conditions.

\section {Results and discussions}
\begin{figure}[htb!]
	\centering
	\includegraphics[width=1.0\columnwidth]{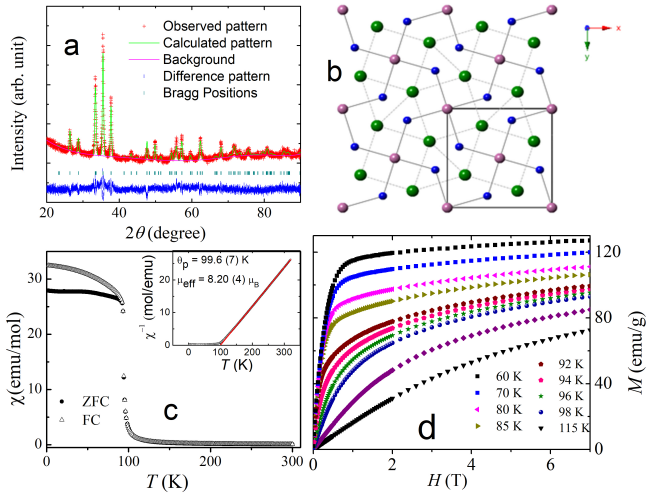}
	\caption{(a-d). (a) Rietveld refinement of powder X-ray diffraction patterns using $P4/mbm$ space group. (b) Crystal structure of Gd$_2$Cu$_2$In is projected along the c-axis. The Gd atoms are represented by green circles where as Cu and In atoms are represented by pink and blue circles respectively. (c) Temperature variation of magnetic susceptibility (FC-ZFC) under the field value of 100 Oe.  and inset picture shows the temperature variation of inverse susceptibility along with Curie-Weiss fitting (red solid line). (d) Isothermal magnetization data for selected temperatures.}
\end{figure}

\subsection{Structural details}

Fig. 1a shows the Rietveld refinement of the XRD pattern and the analysis confirms the single-phase nature of samples with goodness of fit wRp = 4.1 \% and $\chi^2$ = 4.7. The refined lattice parameter values are  a = b = 7.5231 (9) {\AA}  and c = 3.8099  (5) {\AA}. Insert picture of Fig 1 depicts crystal structure of Gd$_2$Cu$_2$In projected along the c-axis. In this structure Gd and Cu respectively occupy $4h$ and $4g$ Wyckoff position whereas In occupies $2a$ site. The structure can be viewed as two alternative layers with one layer containing only Gd atoms and the other layers containing both Cu and In atoms both separated by c/2 (See inset of Fig1). It is interesting to note that the rare-earth atoms form a Shastry-Sutherland atomic-arrangements within Mo$_2$Fe$_2$B type crystal structure. There are two different nearest neighbour for Gd with $d_{Gd-Gd1}$ = 3.6229 (5) {\AA} and $d_{Gd-Gd2}$ = 3.8099 (5) {\AA} and in the basal plane there is an additional nearest neighbour with $d_{Gd-Gd3}$ = 3.9480 (5) {\AA}.  The Gd atoms are surrounded by 6 Cu, 4 In and 3 Gd atoms in their co-ordination shell. The Gd-Cu distances range from 2.9048(3) {\AA} to 2.9650(3) {\AA}  which is significantly larger than the sum of the metallic radii for CN 12 of 3.08 {\AA}. 
\subsection{Magnetic Properties}
Fig 1c. presents the temperature variation of susceptibility in zero-field cooled and field-cooled protocol carried out under the applied field $H$ = 10 mT and the measurement showed a sharp magnetic transition near 94 K indicating presumable ferromagnetic ordering. The Curie temperature is observed to be $\approx$ 94 K from the d$M$/d$T$ vs $T$ plot (picture not shown). Accurate $T_C$ can be estimated from the rigorous critical exponent analysis. Temperature variation of DC susceptibility showed a divergent behavior between FC and ZFC measurements. In general, bifurcation between ZFC and FC arises due to the following reasons (i) short range magnetic ordering and spin glass state (ii) two competing magnetic interactions vis FM/anti-FM interactions which generates substantial magnetic anisotropy field (iii) magnetic inhomogeneity and quenched disorder. Here in the case of Gd$_2$Cu$_2$In we believe that the local anisotropy field or external magnetic field causes the freezing of Gd moments in the direction which is energetically preferred.  Inset plot (b) of Fig1 shows linear behavior of $1/\chi$ vs $T$ data above $T_C$ confirming the Curie-Weiss (CW) behavior. We estimated effective magnetic moment $\mu_{eff}$ to be 8.20 (4) $\mu_B$. The calculated value is greater than theoretical free-ion value expected for Gd3+ spin moment (7.94 $\mu_B$). This excess effective moment could be attributed to the formation of a polarization cloud of the conduction electron around Gd ions. The excess moment arises from Cu 3d orbital which is induced by localized Gd moment. By assuming local Gd moment and ${\mu_{eff}}^2$ = ${\mu_{Ni}}^2$+${\mu_{Gd}}^2$ the excess magnetic moment can be estimated as 1.9 $\mu_B$. The calculated excess moment is close to spin-only moment for Cu$^{2+}$ ion. Fig1d shows set of magnetic isotherms collected in the vicinity of FM-PM transition. At 60 K the magnetic moment was observed to 6.4 $\mu_B$. Fisher et al has observed that the magnetization at 2 K saturates to $\approx$ 7 $\mu_B$/Gd atom both along H$\parallel$c and H $\perp$ c axis, as expected for free Gd$^{3+}$ ions (g $\times$ S = 7$\mu_B$) \cite{Canfield}. Hence it is confirmed that the Cu$^{2+}$ magnetic moment (1.9 $\mu_B$) which is observed from CW fitting arise from the conduction electron polarization and further Cu sub lattice remain non-magnetic in the ordered state. 
\subsection{Magnetocaloric Properties}
In literature, the magnetocaloric effect is characterized by isothermal magnetic entropy change and the $\Delta S_m$ values can be estimated by the following Maxwell's relation \cite{tishin2003magnetocaloric,Bruck}.
\begin{equation}
\Delta S_m(T,H) = \int\limits_{H_i}^{H_f}\big(\frac{\delta M}{\delta T}\big)_HdH^\prime.
\end{equation}
Fig 2a. shows the temperature variation of magnetic entropy change for Gd$_2$Cu$_2$In for different applied fields. Magnetic entropy change (-$\Delta S_m$) shows positive values with single peak shape. The maximum entropy change for 0-7 T field change was 13.8 J/kg.K and even for low field such as 5 T the system exhibits quite large $\Delta S_m$ value of 10.6 J/kg.K. Among the other R$_2$Cu$_2$X (R = Er, Dy, Ho and Gd; X = In and Cd) compounds, Dy, Ho and Er based systems showed larger MCE properties compare to Gd based sample due to multiple phase transitions (See table I).  
Field dependence of MCE has been reported to give additional information regarding the magnetic transition. Within the mean field approximation and for second order magnetic transition Oesterreicher and Parker proposed that the magnetic entropy change follows 
H$^n$ where $n$ is a local exponent taking the value $\approx$ 2/3 near the phase transition \cite{Oesterreicher}. The local exponent ($n$)  can be estimated by fitting the  $ln (\Delta S_m^{pk})_{T=Tc}$ vs $ln H $ plot.  Fig 2(c) shows the field dependence of $\Delta S_m$ at magnetic transition along with the linear fit. The linear fitting shows that the slope is 0.60 (2) indicating slight deviation from the mean field type critical behavior.      
\begin{figure}[htb!]
	\centering
	\includegraphics[width=1.0\columnwidth]{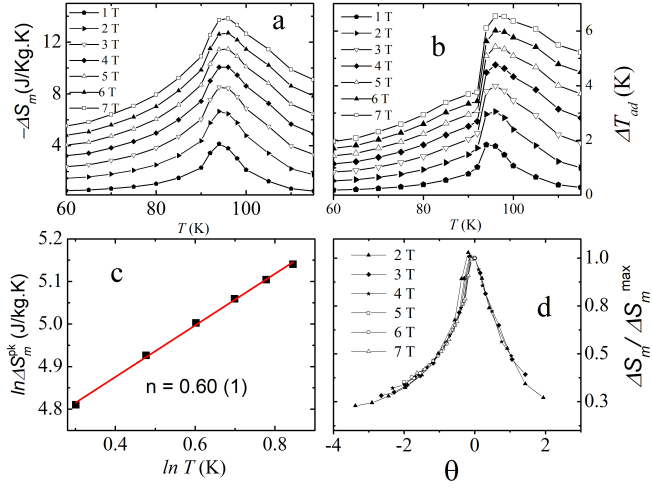}
	\caption{ colour on line (a-d) Temperature variation of - $\Delta S_m$ (a) and $\Delta T_{ad}$ (b) estimated for various applied field using Maxwell's thermodynamic relations. (c) The maximum entropy change as a function of field shown in log-log scale and red line represents the fitted curve. (d) Normalized entropy change as a function of scaled temperature for selected applied field}
\end{figure} 

\begin{figure}[htb!]
	\centering
	\includegraphics[width=1.0\columnwidth]{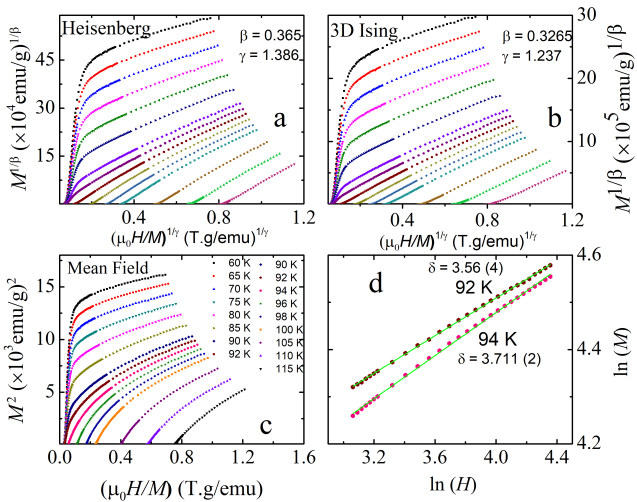}
	\caption{(Colour online) (a and b) represents modified Arrott plots $M^{1/\beta}$ vs $M^{1/\gamma}$  with the critical exponents belonging to Heisenberg model (a) and 3D Ising model (b). (c) Isotherms of $M^2$ vs $H/M$ at selected temperatures near $T_C$. (d) Isothermal magnetization data (critical isotherms) shown in log-log scale and the solid green line represents the linear fit to the data.}
\end{figure}

Adiabatic temperature change ($\Delta T_{ad}$) can be estimated using the Maxwell's relation by the following method \cite{Bruck}; 

\begin{equation}
\Delta T_{ad}(T,H) = - \frac{T\Delta S_M}{C_P(T,H)}.
\end{equation}
We evaluated $\Delta T_{ad}$ using the $\Delta S_m$ curves and zero field heat capacity data. The temperature variation of $\Delta T_{ad}$ resembles the temperature variation of $\Delta S_m$ till 3 T field but we observed the variation of $\Delta T_{ad}$ deviates from smooth peak function for the higher field (Fig2b). However our estimate gives a lower bound value of adiabatic temperature change which is an important parameter for magneto-refrigerant applications and not estimated in this type of materials in previous reports \cite{HoCuIn,GdCuCd,HoCuCd,Zhang_SR}. The maximum value was observed to be  5.4 K and 6.5 K for the applied field 5 T and 7 T respectively. 
\\

Recently, Franco and co-workers have proposed a universal scaling of magnetocaloric effect in order to differentiate between first and second order magnetic transition. According to this model,  for any second order phase transition the normalized entropy change ${\Delta S_M}/{\Delta S_M^{max}}$ (where $\Delta S_M^{max}$ is maximum entropy change)  against scaled temperature ($\theta$) merges into a single curve both below and above the magnetic ordering. In the case of first order magnetic transition the universal curve disperse into branches \cite{franco2008universal,bonilla2010new}. We have used universal scaling method to study the magnetic transition of Gd$_2$Cu$_2$In and the scaling has been done as follows. Rescaling of the temperature axis is done by two reference temperatures viz $T_{r1}$ and $T_{r2}$. 
\begin{equation}
\theta=
\begin{cases}
-(T-T_C)/(T_{r1}-T_C) &  T \leqslant T_C \\
(T-T_C)/(T_{r2}-T_C) &  T>T_C
\end{cases}
\end{equation}
The references temperatures are chosen for each field in such a way that 
\begin{equation}
{\Delta S_m(T_{r1})}/{\Delta S_m^{max}} =  {\Delta S_M(T_{r2})}/{\Delta S_m^{max}} = h (0<h<1) 
\end{equation}
Fig 3d shows the normalized entropy change with respect to scaled temperature constructed by assuming $h$ = 0.5. It is evident from the plot that the normalized entropy collapsed into a single curve for the entire temperature range which is indicating a second order type magnetic transition .
\\
The relative cooling power (RCP) is defined as the amount of heat transfer between the hot and the cold reservoirs in an ideal refrigeration cycle. We estimated relative cooling power by the following expressions. 
\begin{equation}
RCP = \Delta S_m^{max}\times\delta T_{FWHM}.
\end{equation}

In the present investigation, the RCP value for 0-5 T and 0-7 T applied field were 225 J/Kg and 360 J/kg respectively. The magnetocaloric effect for the compound Gd$_2$Cu$_2$In is compared with previous reports on isostructural compounds (See Table1).
\begin{table}
	\caption{Ordering temperature, magnetic entropy change, adiabatic temperature, relative cooling power for Gd$_2$Cu$_2$In are compared with some selected R$_2$Cu$_2$X compounds}
	\begin{tabular}{llllllll}
		\hline
		\hline
		Compound & Transition  & $\Delta S_M^{max}$  &  RCP \\
		& \centering K & J/kg.K &  (J/kg)    \\
		\hline 
		& & &   \\			
		Dy$_2$Cu$_2$In\cite{DyCuIn} & \centering 49.5 & 13.3 &  409   \\
		Ho$_2$Cu$_2$In\cite{HoCuIn} & \centering 30  & 23.6 &  416      \\
		Ho$_2$Au$_2$In\cite{HoCuIn} & \centering21  & 15.1 &  229    \\
		Dy$_2$Cu$_2$Cd\cite{Zhang_SR} & \centering 48  & 13.8 &  381     \\
		Er$_2$Cu$_2$Cd\cite{GdCuCd} & \centering36  & 15.4 &  336     \\
		Gd$_2$Cu$_2$Cd\cite{GdCuCd} & \centering120   & 7.8 &  314    \\
	\bf{	Gd$_2$Cu$_2$In$^*$ }& \bf{\centering 94}  & \bf{11.4} & \bf{ 225}    \\
		
		\hline	
		& & &   \\		
	\end{tabular}
	\noindent
\end{table}
\subsection{Critical behavior and Scaling analysis}
Fig 3c shows the Arrott plots for the compound Gd$_2$Cu$_2$In constructed by plotting $H/M$ Vs $M^2$ which assumes that the critical exponents follows mean field behaviour with $\beta$ = 0.5 and $\gamma$ = 1. The variation of $M^2$ with $H/M$ shows positive slope for full temperature and field range of investigation. According to Banerjee criterion a positive slope in the low field regime of $M^2$ vs $H/M$ is indicative of a second order phase transition \cite{Banerjee}. For mean field type critical behavior, it is expected that the high field magnetization should show set of parallel lines near $T_C$. Further, the critical isotherm at $T$ = $T_C$ should pass through origin if the magnetization can be written as 

\begin{equation}
\mu_0H/M= a(T)+b(T)M^2
\end{equation}
where the $a(T)$, $b(T)$ are temperature dependent constants. From Fig3c it is clear that the magnetic isotherms shows quasi linear behavior $H/M$ Vs $M^2$, critical isotherm does not pass through origin. Hence the mean field model with $\beta$ = 0.5 and $\gamma$ = 1.0 is not suitable exponents for this system. In order to extract the accurate critical exponents for this system we employed commonly used modified Arrott plot technique based on Arrott-Noaks equation of state. According to this model the magnetization (order-parameter) can be written as
\begin{equation}
(H/M)^{1/\gamma} = a(\frac{T - T_{C}}{T}) + bM^{1/\beta}
\end{equation}
where a and b are constants; $\beta$ and $\gamma$ are critical exponents which can be estimated from a iterative non-linear fitting. The details of the scaling hypothesis, renormalization procedure and estimation of critical exponents are given below. 
According to the scaling hypothesis, spontaneous magnetization $M_S(T)$, initial susceptibility and critical isotherms are characterized by set of three critical exponents viz $\beta$, $\gamma$ and $\delta$ respectively. These exponents are strongly inter-related and depends only upon the lattice dimension ($d$) and spin dimensionality ($n_s$). The spontaneous magnetization and initial susceptibility can be expressed as power law expression involving reduced temperature and the critical exponents \cite{fisher1}. 

\begin{equation}
M_{s}(T) = M_{0}(-\epsilon)^{\beta}, \epsilon <  0, T < T_C
\end{equation}
\begin{equation}
\chi_{0}^{-1}(T) =(h_{0}/M_{0})\epsilon^{\gamma}, \epsilon >  0, T > T_C
\end{equation}
\begin{equation}
M = DH^{1/\delta}, \epsilon = 0, T = T_C
\end{equation}

where $\epsilon$ = ($T-T_C/T$) is the reduced temperature;  $M_0$, $h_0$, $m_0$, $D$ are known as critical amplitudes. 
Further, in the critical region, the magnetic equation of state can be written as follows
\begin{equation}
M(H,\epsilon) = \epsilon^\beta f_{\pm}(H/\epsilon^{\beta+\gamma})
\end{equation}
where $f_+$ and $f_-$ are regular analytic functions respectively correspond to below and above $T_C$. 

The magnetization ($M$) and the field ($H$) can be renormalized in such way that the $m \equiv \epsilon ^{-\beta} M(H,\epsilon)$ vs $h \equiv \epsilon ^{-\beta+\gamma} H$ plot collapses into two branches below and above $T_C$. Branching of renormalized magnetic isotherms confirms the validity of the critical exponents and of the universality class.
Estimation of spontaneous magnetization, initial susceptibility and to derive accurate set of critical exponents for a given system are cumbersome process. Initially, modified Arrott plot is constructed by employing equation (7) and by assuming a set of $\beta$ and $\gamma$ value. The spontaneous magnetization and initial susceptibility is estimated by linear extrapolation high field data to the $M^{1/\beta}$ and $(H/M)^{1/\gamma}$ respectively. The $M_S(T)$ vs $T$ and $\chi_{0}^{-1}(T)$ vs $T$ data fitted with equation (8) and (9) to obtain new set of $\beta$ and $\gamma$ values. This procedure is repeated until the critical exponents converge to stable values. We have followed similar procedure to estimate the $\beta$ and $\gamma$, however, the initial values and the university class are guessed in order to avoid too many cycles of constructing MAPs. Fig 3 (a-c) shows both conventional and modified Arrott plots for mean field, 3D Ising and 3D Heisenberg models using the theoretical critical exponents (see Table 2.). It is clear from the plots that 3D Ising and 3D Heisenberg models would be suitable for this systems as the high field data exhibiting set of parallel lines. In order to further narrow down the possible values we have calculated the relative slopes as function of temperature $RS \equiv S(T)/S(T_C)$. The relative slope change should be close to the value 1 for the most suitable modified Arrott plot. The relative slope change for the 3D Ising model showed only 22 \% deviation from the expected values where as mean field and 3D Heisenberg models shows more than 30 \% deviation. Hence we believe that the system could be renormalized into 3D Ising universal class. Franco et al have proposed the local exponent '$n$', which is controlling field dependence magnetic entropy change, can be used to interpret the critical behavior using the following expression \cite{Franco1}. 
\begin{equation}
n = 1 + \frac{\beta-1}{\beta+\gamma}
\end{equation}
using the Widom scaling ($\delta = 1 + \gamma/\beta$) the expression (12) can be re-written as
\begin{equation}
n = 1 + \frac{1}{\delta}(1-\frac{1}{\beta})
\end{equation}
The critical exponent $\delta$ can be found according to the equation (10).  Fig3d shows ln$M$ - ln$H$ plot along with linear fitting and the $\delta$ value was found to be 3.711 (2) for 94 K and 3.56 (4) for 92 K. Using the $n$ and $\delta$ values we derived inital values of $\beta$ and $\gamma$ to construct the modified Arrott plots. Fig4a shows non-linear fitting of spontaneous magnetization and initial susceptibility using the expression (8 and 9). The final fitting gives $\beta$ = 0.312 (2); $T_C$ = 92.97 (4) $T < T_C$ and $\gamma$ = 1.08 (5) $T_C$ = 93.3 $T > T_C$. The estimated $\beta$ value is close to 3D Ising behavior where as $\gamma$ value signifies long range ordering with mean field type behavior. 
Kouvel-Fisher technique is used to deduce the accurate values in literature. According to this method, the power law behavior of the magnetization (Eq.8) and initial susceptibility (Eq.9) can be deduced to simple linear function by dividing the respective first order derivatives. Hence the functions $M_S(dM_S/dT)^{-1}$ and $\chi^{-1}(d\chi^{-1}/dT)^{-1}$ follow linear relation against temperature with slope 1/$\beta$ and 1/$\gamma$. Fig 4c shows $M_S(dM_S/dT)^{-1}$ and $\chi^{-1}(d\chi^{-1}/dT)^{-1}$ vs $T$ plot along with the linear fitting. The estimated exponents are $\beta$ = 0.331 (8); $T_C$ = 93.33 (5) $T < T_C$ and $\gamma$ = 1.034 (6) $T_C$ = 93.54 (1) $T > T_C$ which agrees with estimation from modified Arrott plot technique.
According to static scaling hypothesis, the reliability of the obtained critical exponents can be verified by the universal scaling of the magnetization curves. Using the critical exponents observed from KF method, we constructed  $m \equiv \epsilon ^{-\beta} M(H,\epsilon)$ vs $h \equiv \epsilon ^{-\beta+\gamma} H$ plot (see Fig 4d). It can be clearly seen that all the data collapses into two different curves one corresponds below $T_C$ and other above $T_C$ which is indicating the parameters are reliable and scaling hypothesis valid for the estimated critical exponents.
\begin{figure}[htb!]
	\centering
	\includegraphics[width=1.0\columnwidth]{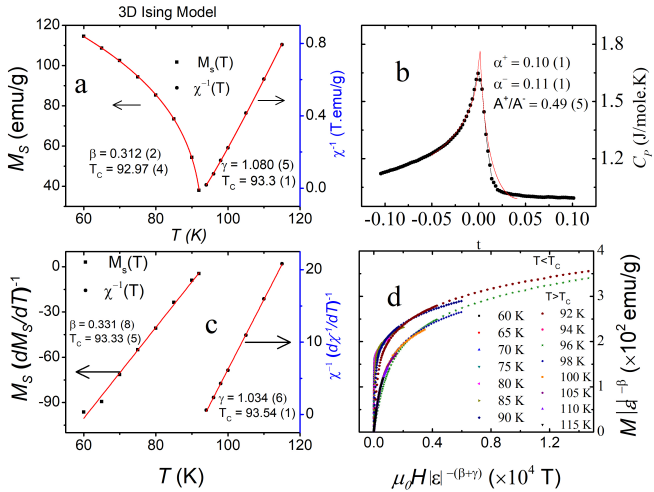}
	\caption{Colour online (a) Temperature variation of spontaneous magnetization and initial susceptibility along with non-linear fit using the expression (8 and 9) (b) Experimental (black circles) and fitted curves (red line) of the specific heat as a function of the reduced temperature in the vicinity of $T_C$ by using equation (14) (c) Kouvel-Fisher plot for the spontaneous magnetization and initial susceptibility [solid red lines represents the linear fit. (d) Scaling plots of $M$$\epsilon^{-\beta}$ vs $H$$\epsilon^{-(\beta+\gamma)}$ below and above $T_C$ using the estimated critical exponents ($\beta$ = 0.331 and $\gamma$ = 1.034)}
\end{figure}
The fourth critical exponent $\alpha$ can be estimated from the heat capacity measurements. Since $\alpha$ can be estimated from the zero field heat capacity, the uncertainty due to the demagnetization field/residual field presence in the magnetization measurements can be avoided. For anti-ferromagnetic systems, $\alpha$ could be more useful parameter than $\beta$ and  $\gamma$ as magnetization is not considered to be the order-parameter. Zero field $C_P$ data near the transition temperature is fitted to the following relation:
\begin{equation}
C_{P} = B + C\epsilon + A^{\pm}|\epsilon|^{-\alpha}(1 + E^{\pm} |\epsilon|^{0.5})
\end{equation}
where $\epsilon$ is the reduced temperature; $A^{\pm}$, B, C and $E^{\pm}$ are adjustable parameters. The relevant information about the critical behavior of the heat capacity and fitting procedures can be found in the following references \cite{Oleaga1,Oleaga2}. The experimental heat capacity data along with fitting curves are shown in Fig4b. The value of critical exponent was observed to be $\alpha^+$ = 0.10 (1) and $\alpha^-$ = 0.11 (1); the ratio of the critical co-efficients is $A^+/A^-$ = 0.49 (5). Both the critical exponent and the ratio values confirm the 3D-Ising universality class which is in concert with our MAP, KF analysis. The observed values are far away from 3D Heisenberg and XY universality which is in concert with our estimation from magnetization measurements.

\begin{table}
	\caption{Estimated critical exponents ($\beta \gamma \delta \alpha$) for Gd$_2$Cu$_2$In compared with theoretical critical exponents }
	\begin{tabular}{ccccccccc}
		\hline
		\hline
		Material& Technique & $\beta$ & $\gamma$ &\centering $\delta$ & $\alpha$ & $A^{+}/A^{-}$ \\
		\hline
		Gd$_2$Cu$_2$In & MAP & \centering$0.3147$& \centering$1.095$& \centering$4.48$ & & \\
		
		& KF &\centering$0.312 $&\centering$1.106 $&\centering$4.54 $ & & \\

		& CI & & &\centering$4.545 $ & & \\

		& $C_P$ & & & & $0.11$ & $0.49 $\\
		\hline	
		& Mean-Field & \centering0.5 & \centering1.0 &\centering3.0 & 0 & - \\
		Theory   & 3D-Ising & \centering0.365 & \centering1.336 & \centering4.8 & 0.11 & 0.524 \\
		& Heisenburg & \centering0.325 & \centering1.24 & \centering4.82 &-0.115 & 1.521 \\
		
		\hline	       
	\end{tabular}
\end{table}

All the critical exponents estimated from various techniques along with the known theoretical model values are summarized in table II. As we mentioned earlier that the $\beta$ and $\alpha$ value indicate the systems possess 3D Ising type behavior where as $\gamma$ and $\delta$ values are located in between 3D Ising and mean field value. Our observations indicate a possible crossover between 3D-Ising to mean field behavior in the vicinity of the phase transition. Crossover behavior often observed in many hole doped RMnO$_3$ (R = Lanthanide) manganites. Kim et al have observed in La$_{0.75}$Sr$_{0.25}$MnO$_3$ sample that the critical exponents lie in between 3D Ising model vales and mean field values \cite{LSMO}. Critical behavior of CrSiTe$_3$ showed cross over behavior where the system showed 2D Ising model value coupled with long-range spin interactions \cite{CrSiTe}. Recently, Dey et al. have observed cross over between 3 D Heisenberg behavior to mean field behavior for the system MnCr$_2$O$_4$ \cite{Dey}. It is important to understand the cross over behavior in terms of length and temperature scales. The correlation length $\xi$ = $\xi_0$ $|\epsilon|^{-\upsilon}$, where $\upsilon$ is the correlation length exponent, can be roughly estimated by using the relation $\upsilon$ = $(2-\alpha)/d$ \cite{LSMO}; By assuming typical values for $\xi_0$ to be 5-10 {\AA} we estimated correlated length to be 51 {\AA} to 102 {\AA} for $\epsilon$ $\approx$ 0.025. This indicates that the correlation length either comparable or much smaller than the typical grain size of a polycrystalline materials. Hence the critical fluctuations is more dominant than the mean field behaviour to study the cross over behaviour. Our $C_P$ and magnetization studies are within this regime of uncertainty hence either neutron diffraction studies or single crystal magnetization measurements needed for the studying the complete cross over behavior. 
Finally, it is clear from the earlier discussions the system does not complete belong to 3D short range interaction models and the estimated exponents are in between the 3D Ising and mean field values. The extended type of interaction can be represented by $J(r)$ $\approx$ $r^{-(d+\sigma)}$ where $r$ is the distance and $\sigma$ is the range of interaction \cite{GdNiAl2}. For a homogeneous magnet, the susceptibility exponent $\gamma$ can be expressed by the following relation\cite{Fisher,Nair} 

\begin{equation}
\begin{split}
\gamma = 1 + \frac{4}{d}\bigg(\frac{n_S+2}{n_S+8}\bigg)\Delta \sigma+\frac{8(n_S+2)(n_S-4)}{d^2(n_S+8)^2} \\
\times\bigg(\frac{2G(d/2)(7n_S+20)}{(n_S-4)(n_S+8)}\bigg)\Delta \sigma^2
\end{split}
\end{equation}
where $\Delta \sigma$ = $\sigma - \frac{d}{2}$; $G(d/2) = 3 - \frac{1}{4} (\frac{d}{2})^2$. In literature, it is customary to solve this equation counter intuitively by choosing correct set of parameter $(d:n)$ in order to find the close $\gamma$ value for the system. In our case since $d$ = 3; and $\gamma$ $\approx$ 1, we believe that the trivial solution for the equation (15) would be $\Delta \sigma$ = 0 which gives $\gamma$ value $\approx$ 1 .This assumption is not completely unphysical because in the renormalization group theory analysis, $\sigma$ = 1.5 signifies mean field type behavior with long range interaction where as $\sigma$ $\approx$ 2 implies a possible short range correlation \cite{GdNiAl2,CrSiTe,Nair}. The critical exponents are inter-connected by the following relations\cite{CrSiTe}. $\upsilon$ = $\gamma/\sigma$; $\alpha$ = 2-$\upsilon d$; $\beta$ = (2-$\alpha$-$\gamma$)/2; $\delta = 1 + \gamma/\beta$. By using the $\alpha$ and $\gamma$ value, accurate $\sigma$ was estimated to be 1.587 which lead to spin interactions $J(r)$ decays as $r^{-4.587}$ expected for long range interactions.

\section{Summary and Conclusions}
In conclusion, we observed large magnetocaloric effect in Gd$_2$Cu$_2$In  compound in the vicinity of magnetic transition. For the field value of 0-5 T and 0-7 T the $\Delta S_m$ value was observed to be 10.6 J/kg.K and 13.8 J/kg.K respectively. We presented a comprehensive and details studies of the critical behavior for the compound Gd$_2$Cu$_2$In  near the transition temperature. The PM-FM phase transition is observed to be second order in nature. Estimation of four critical exponents using various analysis (MAP, KF, critical isotherm and $C_P$ critical exponent studies) show that the system belong to 3D Ising type behaviour coupled with long range interaction. The spontaneous magnetization exponent ($\beta$) and heat capacity exponent $\alpha$ values close to the 3D Ising behaviour where as other two exponents indicate a long range or mean field type interactions. We believe the system magnetically orders below 94 K where Gd 3+ moments arranged uniaxially along long tetragonal axis 'c'. Uniaxial arrangment of Gd moments suggest strong magnetocrystalline anisotropy presence which may be a plausible reason for the observed bifurcation between FC-ZFC magnetization. Finally, the extended range of interaction $J(r)$ decays as $r^{-4.587}$ expected for long range interaction. The cross over from 3D Ising to mean field type behavior evident from the magnetization and heat capacity critical exponents.

\section{Acknowledgments}
RKK acknowledges the FRC/URC Postdoctoral fellowship. AMS thanks the SA-NRF (93549) and the FRC/URC of UJ for financial assistance. \\
$^*$KRK and HSN contributed equally to this work. 


\end{document}